\documentclass[aps,prb,amsmath,twocolumn,amssymb,titlepage,reprint]{revtex4-2}

\usepackage{graphicx}
\usepackage{siunitx}
\usepackage{xcolor}
\usepackage{amsmath}
\usepackage[english]{babel}
\usepackage[colorlinks=true]{hyperref}
\usepackage{lipsum}
\usepackage{float}
\usepackage{placeins} 
\definecolor{ccol}{rgb}{0,0.59,0.51}
\definecolor{ucol}{rgb}{.27,.39,.67}
\definecolor{red}{cmyk}{0.25,1,1,0}
\hypersetup{citecolor=ccol, urlcolor=ucol}
\usepackage[T1]{fontenc}
\usepackage{todonotes}
\usepackage{appendix}

\usepackage{soul}
\usepackage{comment}
\usepackage{verbatim}

\begin{document}

	\title{Efficient fiber coupling of telecom single-photons from circular Bragg gratings}

	\author{Nam Tran$^{1}$}
	\thanks{These three authors contributed equally}
	\author{Pavel Ruchka$^{2}$}
	\thanks{These three authors contributed equally}
    \author{Sara Jakovljevic$^{2}$}
    \thanks{These three authors contributed equally}
    \author{Benjamin Breiholz$^{1}$}
    \author{Peter Gierß$^{1}$}
    \author{Ponraj Vijayan$^{1}$}
    \author{Carlos Eduardo Jimenez$^{3}$}
    \author{Alois Herkommer$^{3}$}
    \author{Michael Jetter$^{1}$}
    \author{Simone Luca Portalupi$^{1}$}
    \author{Harald Giessen$^{2}$}
    \author{Peter Michler$^{1}$}
	\affiliation{$^1$ Institut für Halbleiteroptik und Funktionelle Grenzflächen, Center       for Integrated Quantum
		Science and Technology (IQ\textsuperscript{ST}) and SCoPE, University of Stuttgart, Allmandring 3, 70569 Stuttgart, Germany \\
        $^2$ 4. Physikalisches Institut,  Center for Integrated Quantum
		Science and Technology (IQ\textsuperscript{ST}) and SCoPE, University of Stuttgart,
        Pfaffenwaldring 57, 70569 Stuttgart, Germany,
        $^3$ Institute of Applied Optics (ITO) and Research Center SCoPE, University of Stuttgart, Pfaffenwaldring 9, 70569 Stuttgart, Germany}

	\begin{abstract}
        Deterministic sources of quantum light are becoming increasingly relevant in the development of quantum communication, particularly in deployed fiber networks.  Therefore, efficient fiber-coupled sources at telecom wavelength are highly sought after. With this goal in mind, we systematically investigate the fiber coupling performance of quantum dots in optical resonators under three experimental configurations.
        We quantify coupling efficiency and sensitivity to spatial displacement for single-mode fibers with 3D printed optics on their tip, and benchmark their behavior over a commercial cleaved-cut fiber and a standard optical setup.
        The reduction of the required optical elements when operating with a lensed or a bare fiber allows for an increased end-to-end efficiency by a factor of up to $3.0\pm0.2$ over a standard setup. For the perspective of realizing a mechanically stable fiber-coupled source, we precisely quantify the spatial tolerance to fiber-cavity misalignment, observing less than $\SI{50}{\percent}$ count rate drop for several micrometers displacement. These results will play a key role in the future development of fiber-coupled sources of quantum light.

	\end{abstract}
	
	\maketitle


	\section{Introduction}
	Single-photon emitters are key building blocks for future implementations in quantum communication \cite{O'brien2009, wehner2018quantum, heindel2023quantum, vajner2022quantum}. Solid-state emitters such as semiconductor quantum dots (QD) have shown great results in terms of their optical properties in the last decades \cite{michler2024semiconductor, senellart2017high}. This pushed the development of QD sources with high indistinguishability \cite{somaschi2016near}, high brightness \cite{tomm2021bright, ding2025high}, and high single-photon purity \cite{SchweickertPurity2018, hanschke2018quantum}. The challange posed by the photon extraction from the solid-state matrix \cite{barnes2002solid} has been overcome by integrating microcavities reaching end-to-end efficiencies of $\SI{57}{\percent}$ \cite{tomm2021bright} and $\SI{71.2}{\percent}$ \cite{ding2025high}. Circular Bragg gratings (CBG) have gained particularly interest over the past years since their far-field cavity mode exhibits a Gaussian profile, and they provide broadband extraction efficiency, resulting in high detected count rates \cite{sapienza2015nanoscaleCBG,liu2019solidCBG, WangHuiCBG, NawrathCBGTelecom}. As the current communication technology is built upon a global fiber network, the wavelength of the photons must be in the telecom O- ($\SI{1310}{\nano\meter}$) or C-Band ($\SI{1550}{\nano\meter}$) due to the low wavepacket dispersion and low loss in optical glass fibers at these wavelengths.\\
    In recent years, QDs in the telecom-range have been successfully grown and have exhibited promising results \cite{yu2023telecom}. Also for telecom QDs, their performances have been improved via the integration into microcavity structures such as CBGs \cite{NawrathCBGTelecom, JoosCBGTelecom, holewaCBGTelecom}, photonic crystals \cite{PhilippsPCTelecom}, and open-fiber cavities \cite{MaischGrammelOpenFPTelecom}. Usually, the photons are generated and collected through a microscope objective in a free space configuration. However, to access a fiber network and perform certain protocols, the photons have to be coupled into a single-mode fiber from free space. This process introduces unwanted losses, reducing the total amount of collected photons. Specialized tailored single-photon sources that can mitigate these photon losses are thus required. This becomes even more significant with the growing demand of highly efficienct single-photon sources. \cite{heindel2023quantum, neuwirth2021quantum, thomas2021race}.
    \\
    Efficient fiber coupling can be improved by placing a fiber above the QD to directly couple single-photons without the need of a microscope objective. In this way, the amount of required optics can be minimized, resulting in enhanced fiber-coupling efficiency.
    Additionally, with this approach, the fiber can be attached or glued to the sample, allowing plug-and-play operation \cite{BremerKsenia, Rickert, LRickert2025, RickertMicropillars, schlehahn2018stand}. This provides a more practical and user-friendly fiber-coupled single-photon source where free-space alignment is not needed. 
    Such advantages could be in particular promising for the development of compact, transportable single-photon sources \cite{TGaoOutOfLab}. The combination of the monolithic integration of fiber and CBG and efficient fiber-coupling will play a key role in the future.
    \\
    In this work, we perform a quantitative study on the achievable coupling efficiencies by investigating the influence of spatial lateral and vertical displacement of the fiber over CBG. This knowledge on the permitted displacement and optimal positioning of the fiber with respect to CBGs is crucial for the future monolithic integration of fiber and CBG. We compare three different coupling setups using a fiber with a 3D-printed lens \cite{Ruchka_2022} at the fiber tip, a bare fiber with a cleaved end, and a commercial low-temperature objective as a reference. Furthermore, we investigate the reproducibility of the coupling by performing the measurements on four different CBGs for each of the setups. Our findings provide important foundations for further miniaturization of fiber-coupled single-photon sources in the telecom C-band.

	
	\section{Results and Discussion}
    \subsection{\label{sec:Fabrication}Device fabrication}
    
    We design and fabricate on {a single-mode fiber tip a 3D-printed aspheric lens with an NA of $0.6$. To achieve such an NA, we first splice a piece of no-core fiber with a length of $\SI{550}{\micro\meter}$ directly onto the single-mode fiber. The lens is then 3D-printed via two-photon polymerization. 
    \\
    For the single-photon source, semiconductor quantum dots were grown by metal-organic vapor-phase epitaxy (MOVPE) in the Stranski-Krastanov growth mode \cite{yamaguchi2000stranski} based on an InAs material including a metamorphic buffer layer \cite{SittigMMB, MMBSemenova} on a GaAs substrate. The fabrication of CBGs embedding QDs is realized by a flip-chip process onto a silicon carrier and a series of processing steps comprising wet chemical etching, electron beam lithography, and ICP-RIE dry etching \cite{KolatschekCBG-OBand}. With an additional gold back reflector, optimal out-coupling and Gaussian far-field can be achieved \cite{NawrathCBGTelecom, JoosCBGTelecom}. For more information about the fabrication of the 3D-printed lens on fiber and CBGs, we refer to the SI.
    \\
	As one of our goals was to investigate the reproducibility of the coupling, we performed measurements on four different CBGs, each in the following configurations as sketched in Fig.\ref{fig1}: (a) lensed fiber (NA=0.6), (b) SMF-28 bare fiber ($\text{NA}=0.14$), and (c) microscope objective ($\text{NA}=0.82$). The SMF-28 bare fiber is used as a direct reference to the lensed fiber to investigate the behavior of the 3D-printed lens. For each configuration, we investigated the capability of the devices of (i) imaging the sample surface, (ii) sensitivity of the coupling for spatial displacement in X-,Y-, and Z-direction, and (iii) the maximum coupling of the single-photons emitted from the CBG into the fibers. 
    \\
        \begin{figure}
		\includegraphics[width=\columnwidth]{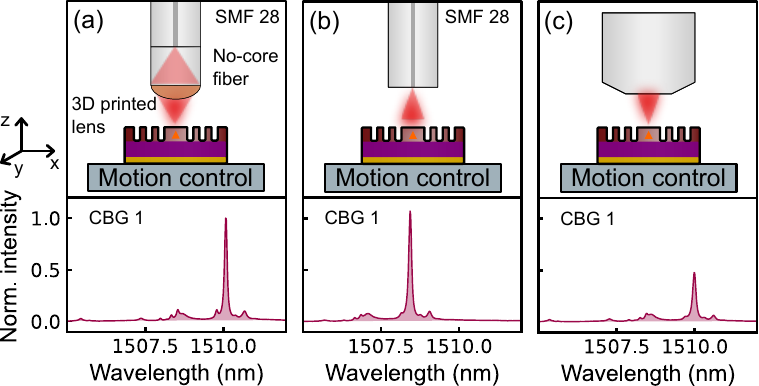}
		\caption{Experimental setup for the three configurations: (a) 3D-printed lensed fiber with a design numerical aperture of $\text{NA}=0.6$ fabricated via two-photon polymerization. (b) Cleaved SMF-28 bare fiber. (c) Microscope objective with nominal $\text{NA}=0.82$. In the respective bottom panels of each configuration, the measured µ-PL QD spectrum of CBG 1 (see Tab.\ref{tab:Table1}) is presented where each spectra is normalized to the overall maximum. The blue-shift in wavelength in (b) is induced by strain due to a slight detachment of the semiconductor membrane. }
		\label{fig1}
	\end{figure}
    To make an accurate investigation of these performances in the three different configurations, the same CBGs have to be directly compared. Exemplary, in the bottom panels of Fig.\ref{fig1}, the µ-photoluminescence (µ-PL) spectra of the QD within CBG 1 (see Tab.\ref{tab:Table1}), measured in the three configurations, are presented. In total, four different CBGs were investigated for statistical purposes showing that the results are reproducible. Here, non-resonant excitation was applied where the brightest line was used for all measurements. While the spectral features are the same in the three cases, a blue-shift of \SI{1.6}{\nano\meter} was observed in the case of the bare fiber. This was observed consistently for all emitters and attributed to a detachment of the semiconductor from the sample carrier after several cooling cycles. This led to a change of the strain conditions on the QD and therefore to a blue shift of the spectra. Despite that, the emitter performance remained comparable. A more detailed discussion on the achievable count rate in each configuration will be given in Sec. \ref{sec:Brightness}.

    \subsection{\label{sec:Imaging}Imaging of sample surface}
    To make a direct comparison of the same CBGs with the three configurations, it is essential to resolve the surface of the sample to retrieve the same cavity. With the microscope objective it is possible to directly project the image of the sample surface onto a camera when illuminating with a white light source, hence utilizing a free-space configuration. When using a structure consisting of only one single-mode fiber, imaging the surface becomes more challenging and cannot be performed in a similar way to a microscope objective as only one spatial mode can propagate within the fiber. As a consequence, most of the spatial information of the sample surface would get lost. Therefore, other methods have to be applied to image the sample surface.
	For the following experiments we made use of a high stability bath cryostat (Attocube liquid 100) where the relative fiber-sample position can be controlled by precise piezo-scanners (Attocube ANS100 series). With additional interferometric displacement sensors in the lateral XY-direction (Attocube FPS3010), the position is reliably controlled with nanometer precision. Moreover, with additional coarse steppers (Attocube ANP101 series) we are able to navigate around the entire sample to target the desired CBGs.
    \begin{figure}
		\includegraphics[width=\columnwidth]{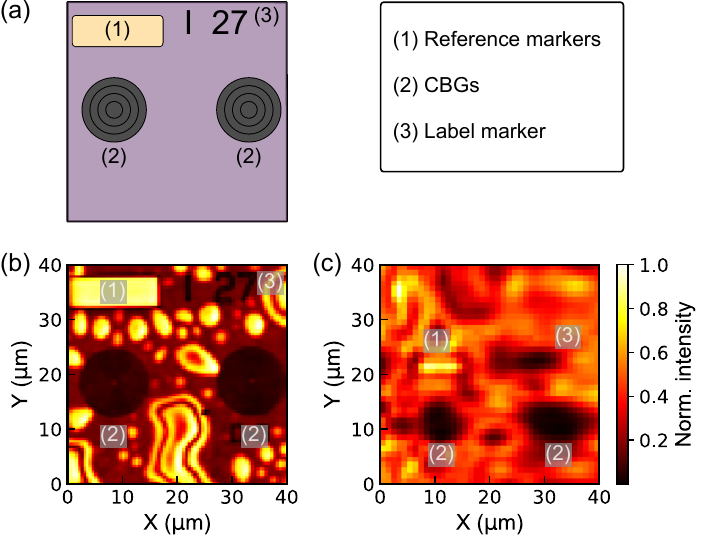}
		\caption{(a) schematic illustration of the area on the sample which was scanned. (b), (c) $\SI{40}{\micro\meter} \times \SI{40}{\micro\meter}$ XY-reflectivity scan on the sample surface utilizing the (b) lensed fiber and a (c) bare fiber. A red laser ($\SI{658}{\nano\meter}$) is injected into the fibers while the sample is scanned  in the lateral XY-direction. Simultaneously, the reflected signal of the laser ($\SI{90}{\percent}$) is collected via a 90:10 fiber beam splitter resulting in an intensity contrast map for the lensed fiber and bare fiber. The right image depicts a slightly vertically shifted region compared to the left image.}
		\label{fig2}
	\end{figure}
    Imaging the sample surface with the fiber structure is then performed in a reflectivity measurement. Via a 90:10 fiber beam splitter, a red laser ($\SI{658}{\nano\meter}$) is inserted ($\SI{10}{\percent}$) into the fiber. While scanning the sample in the lateral XY-direction, the reflected laser signal ($\SI{90}{\percent}$) is detected. Various structures on the sample, which are schematically depicted in Fig.\ref{fig2} (a), such as reference markers, CBGs, and label markers reflect, scatter, or absorb the laser light differently (see SI). Fig.\ref{fig2} (b) displays the measured reflectivity map for which illumination and collection happen through the same lensed fiber (configuration as in Fig.\ref{fig1} (a)). Fig.\ref{fig2} (c) depicts the same measurement when a bare fiber is employed (configuration as in Fig.\ref{fig1} (b)). The size of both scans is set to $\SI{40}{\micro\meter} \times \SI{40}{\micro\meter}$. With the lensed fiber, the resolution is high enough to resolve the structures. The distance between the CBGs were designed to be $\SI{25}{\micro\meter}$ which can be also extracted from the scan in Fig.\ref{fig2} (b), indicating that the lensed fiber can faithfully image the sample surface. In comparison, when using the bare fiber, the same structures are also visible, yet not resolved well enough to identify the labes. The comparison between the two scans prove that the lensed fiber allows for an accurate visualization of the sample surface: this becomes very important when specific photonic structures need to be employed in the experiment. Although more challenging and time-consuming, the same CBGs measured with the lensed fiber were found back with the bare fiber. This was necessary to provide a quantitative comparison among the three experimental configurations.
    \\
    In the future, when other photonic structures are targeted, the lensed fiber could also be appealing since the large spatial resolution would be advantageous. The spatial resolution of the lensed fiber is estimated to be around $\SI{600}{\nano\meter}$ when assuming the parameters in our experiments as $\lambda=\SI{658}{\nano\meter}$ and $\mathrm{NA}=0.6$. This resolution is high enough to resolve most of the photonic structures used in photonic integrated circuits \cite{wang2020integrated, Harris2014, Goyvaerts:20} such as waveguides, beam-splitters \cite{schwartz2018fully, hepp2020purcell, hornung2024highly}, or other micro-resonators \cite{vahala2003optical}. With the bare fiber, the resolution is limited by the mode field diameter (MFD) of the fiber which is mostly in the range of micrometers. The spatial resolution could be improved by deploying a ultra-high-NA (UHNA) fibers exhibiting smallers MFDs, however, still in the micrometer range. To gain the lowest spatial resolution, the bare fiber has to be operated with the smallest possible fiber-sample distance as the beam profile diverges with larger distance, ultimately reducing the resolution. This could pose potential risks of damaging the structures on the sample. In contrast to this, with a working distance of $\SI{36}{\micro\meter}$ for the lensed fiber, the fiber-sample distance is large enough to safely operate. The lensed fiber shows a clear advantage over the bare fiber in terms of navigating and resolving structures on the sample reducing at the end the time-consumption.

            	\begin{figure*}[htbp!]
		\includegraphics[width=\textwidth]{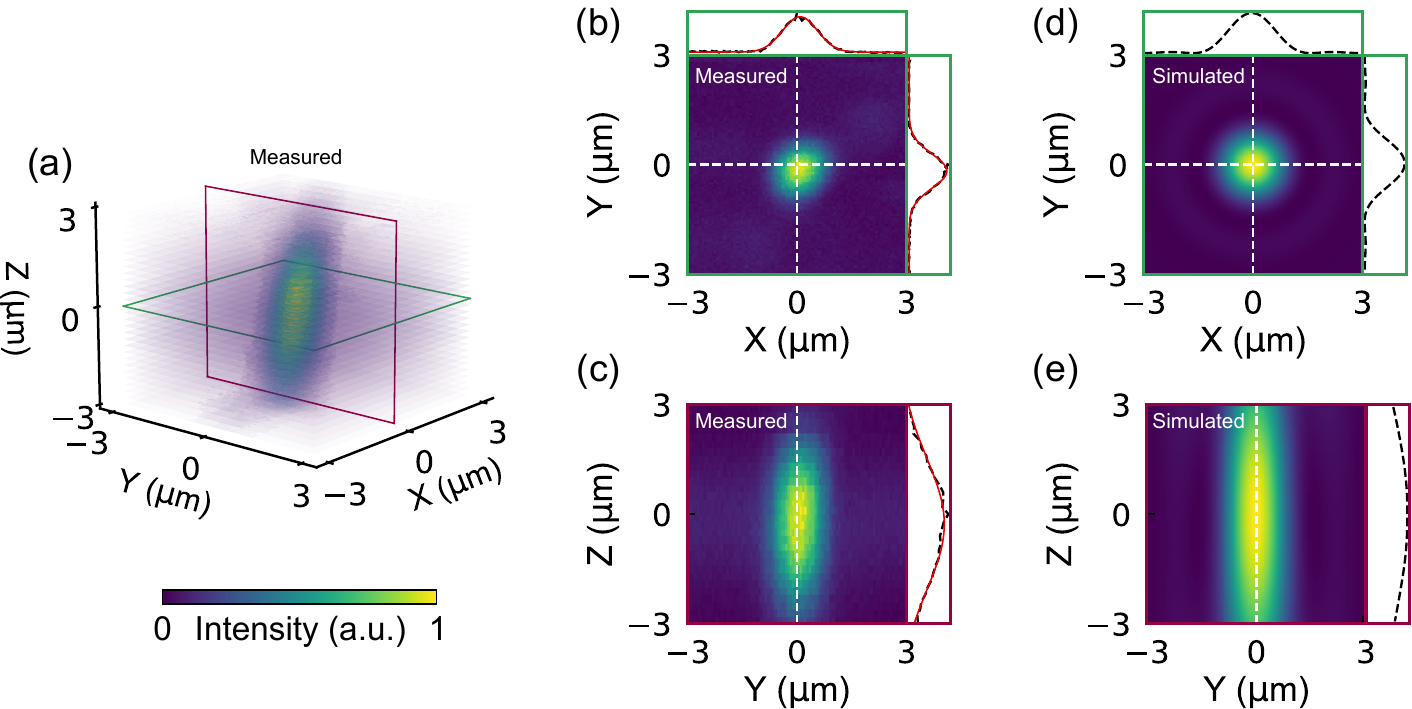}
		\caption{(a) XYZ-scan of the lensed fiber obtained by exciting the QD in p-shell excitation. (b) The XY-scan corresponding to the focal spot is shown and indicated by the green solid line in (a). The white dashed lines depict the cut through the maximum in X- and Y-direction resulting in an average FWHM\textsubscript{XY} in the lateral direction of $\SI{1.22\pm0.2}{\micro\meter}$. (c) Cut in the YZ-direction extracted from the cut of the dark red line from (a). The average FWHM\textsubscript{YZ} is determined to be $\SI{4.56\pm0.7}{\micro\meter}$. (d) Simulation in the XY-plane of the 3D printed lens on fiber yielding FWHM\textsubscript{XY,sim} of $\SI{1.44}{\micro\meter}$. (e) Simulation in the YZ-plane resulting in FWHM\textsubscript{YZ,sim} of $\SI{8.10}{\micro\meter}$}
		\label{fig3}
	\end{figure*} 
        \begin{figure*}[htbp!]
		\includegraphics[width=15cm]{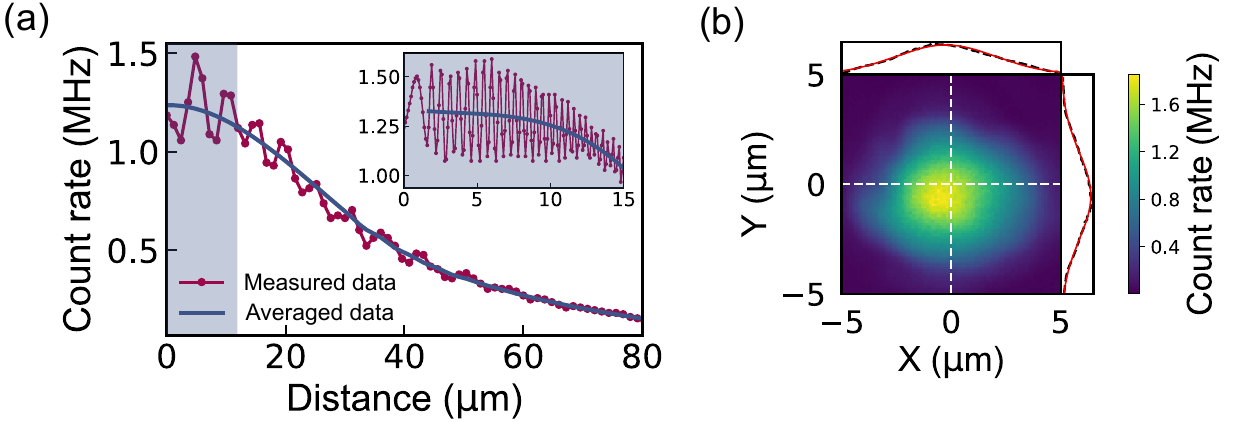}
		\caption{(a) Measured count rate over the fiber-sample distance using the bare fiber. In the inset, a finer scan from the grey shaded area can be observed. The dark red and blue line correspond to the measured data and averaged data. (b) An XY-scan at the last observable oscillation maximum before contact with the sample is shown with an average FWHM\textsubscript{XY} of $\SI{4.90\pm0.34}{\micro\meter}$.}
		\label{fig4}
	\end{figure*} 
    \subsection{\label{sec:PSF}Spatial sensitivity}
	Besides of being able to navigate to the desired CBG, it is necessary to reliably couple the photons emitted from the CBG into the fiber. For this purpose, the fiber must be spatially aligned to the far-field of the cavity mode. Hence, the behavior of the count rate in dependence of the spatial lateral XY- and vertical Z-direction is investigated and quantified. Understanding the spatial sensitivity of the fiber becomes a major aspect in terms of feasibility for fiber-coupled single-photon sources \cite{Schwab:22}. 
	To determine the spatial tolerance of the lensed fiber, a µ-PL map in the XY-direction is scanned around the CBG. While exciting the QD via either cw p-shell ($\lambda_{exc} = \SI{1512}{\nano\meter}$) or cw and pulsed non-resonant pumping ($\lambda_{exc} = \SI{780}{\nano\meter}$), the emitted light is collected and sent to superconducting nanowire single-photon detectors (SNSPD - Single Quantum). This process is repeated for different positions in Z-direction to also probe the sensitivity in the vertical direction. Here, the Z-position is changed step-wise with a stepping size of $\SI{200}{\nano\meter}$ over a range of $\SI{6}{\micro\meter}$ symmetrically around the focal spot of the lensed fiber. The resulting XYZ-scan is depicted in Fig.\ref{fig3} (a) where the green and red solid lines indicate the cut in the (b) lateral XY-plane of the focal spot and (c) vertical YZ-plane, respectively. These measurements are commonly referred to as the Point Spread Function (PSF) \cite{cox2006optical,NWANESHIUDU20121} and were also performed with the microscope objective as a reference (see SI). In Fig.\ref{fig3} (b), a Gaussian fit was applied to extract the distance at which the maximum signal drops to half of the maximum intensity (full width at half maximum \textit{FWHM}) in X- and Y-direction indicated as the white dashed lines. The same procedure was also performed for the vertical YZ-Plane (Fig.\ref{fig3} (c)). Multiple measurements of the spatial sensitivity were performed on different CBGs. In Table \ref{tab:Table1}, the arithmetic mean value and respective standard deviation of all measured FWHM in XY, and YZ are listed. For the sake of clarity, the FWHM of XY corresponds to the mean value of the FWHM in X- and Y-direction. The average experimental value correspond to $\text{FWHM}_{\text{XY}} = \SI{1.22\pm0.2}{\micro\meter}$ and $\text{FWHM}_{\text{YZ}} = \SI{4.56\pm0.70}{\micro\meter}$. From simulations (see SI) of the XY-plane (Fig.\ref{fig3} (d)) and YZ-plane (Fig.\ref{fig3} (e)), the theoretical value $\text{FWHM}_{\text{XY,sim}} = \SI{1.44}{\micro\meter}$ is in good agreeement with the experimental value, whereas $\text{FWHM}_{\text{YZ,sim}} = \SI{8.10}{\micro\meter}$ is slightly larger to the experiment. In our simulations, we assume a Gaussian wave propagating from the interface between the single-mode fiber and the no-core piece to the 3D-printed lens (see SI). However, in the experiment, the light is collected from a CBG emitting in a certain NA which could explain the deviation between simulation and experiment in the vertical direction.
    \\
	As a comparison to the lensed fiber, the spatial sensitivity of the bare fiber was also analyzed. To analyze the behavior of the coupling in the vertical direction, the count rate in pulsed non-resonant excitation ($\lambda_{exc} = \SI{780}{\nano\meter}$) was directly measured while changing the distance of the bare fiber to the the sample. With coarse steps, the distance is step-wise changed by $\SI{1.2\pm0.2}{\micro\meter}$ over a range of $\SI{85}{\micro\meter}$. The according results in Fig.\ref{fig4} (a) indicate a continuous signal drop with a superimposed oscillation in the count rate over the entire range. These oscillations are further resolved in the inset using a fine step size covering a range of $\SI{15}{\micro\meter}$ in steps of $\SI{100}{\nano\meter}$. 
    The reason for the oscillations are interference effects arising from a Fabry-Pérot (FP) cavity formed between the bottom gold mirror of the sample and the cleaved fiber tip. With the measured oscillations and expected resonance condition $\lambda/2$ for a FP-cavity, the position in the vertical Z-direction could be reliably measured. This oscillating behavior has been also simulated for circular Bragg gratings \cite{Rickert} and measured on a QD-micropillar \cite{RickertMicropillars}. In the inset of Fig.\ref{fig3} (a), the first oscillation at a distance of $\sim \SI{0}{\micro\meter}$ exhibits a longer oscillation period than the other oscillations due to the fiber tip coming into contact with the sample surface. As the fiber may be slightly tilted, estimated to be below $\SI{2}{\degree}$, the sample can still be moved even if the fiber tip touches the sample.  Looking further into the coupling, a plateau of the count rate between $0$ to $\SI{10}{\micro\meter}$ is observed before a continuous drop occurs, indicated with the blue averaged line. Within this region, the bare fiber ideally acts as FP-cavity while with larger distance the cavity becomes more unstable, resulting in a drop of oscillation amplitude. At distances larger than the plateau, the beam diameter of the emitted light from the CBG expands larger than the mode field diameter (MFD $\approx \SI{10.4\pm0.5}{\micro\meter}$) of the bare fiber, ultimately impacting the in-coupling efficiency. Using ray optics, the distance $L$, at which the beam profile of the CBG becomes larger than the MFD of the fiber, can be estimated with $L=\text{MFD}/(2\cdot \text{NA}_{\text{CBG}})=\SI{13.0\pm0.6}{\micro\meter}$ where NA\textsubscript{CBG} corresponds to the numerical aperture of the CBG simulated to be around 0.40 (see SI). This value is in in the range of the measured $\SI{10}{\micro\meter}$. In a sense of practical use, the plateau can be defined as the working range (WR) of the bare fiber. To obtain a similar comparison to the lensed fiber, the FWHM\textsubscript{Z} for the bare fiber is determined to be at around $\SI{46}{\micro\meter}$ calculated from the averaged line. Furthermore, an XY-scan was performed at the closest possible distance, where the last maximum of the count rate is found,  before the fiber tip touches the sample, resulting in the scan displayed in Fig.\ref{fig4} (b). 
    From multiple measurements of such scans, the average FWHM\textsubscript{XY} of the bare fiber corresponds to $\SI{4.90\pm0.34}{\micro\meter}$. It would be expected that the the measured FWMH\textsubscript{XY} is in the same range as the MFD of the fiber. Since the MFD is defined as the distance where the mode intensity cross section within the fiber drops to $1/\mathrm{e}^2$, we use the FWHM of the mode intensity of $\SI{6.14\pm0.3}{\micro\meter}$ to make a comparison to our measured value. The measured value seems to be lower from the expected value which could be again attributed to the positional tilt as mentioned before. In terms of the overall spatial sensitivity, the bare fiber exhibits larger FWHMs in both the lateral and vertical direction in comparison to the lensed fiber. 
    \\
    Even though the larger FWHMs of the bare fiber might seem more advantageous, the fiber-sample distance must be set to a maximum of the oscillations, demanding precisions of tens of nanometers. The contrast of the maximum ($\SI{1.58}{\mega\hertz}$) and the minimum ($\SI{1.07}{\mega\hertz}$) of the oscillations amounts to $\SI{67}{\percent}$. It is therefore paramount to be placed at a maximum to maintain the optimal efficiency of the device. Alternatively, by applying an anti-reflection coating onto the fiber tip, the oscillations could also be avoided \cite{Ristok:22}. On the other hand, the lensed fiber exhibits excellent stability in terms of the in-coupling of the single-photons once the fiber-sample distance is adjusted.

            	\begin{table*}[htbp!]
		\centering
		\begin{tabular}{c|c|c|c|c|c}
			& & \textbf{Lensed fiber} & \textbf{Bare fiber} & \textbf{Microscope objective} & \textbf{Wavelength (nm)} \\
			\hline
			FWHM & XY (\(\mu\)m) & $1.22\pm0.20$ & $4.90\pm0.34$ & $1.69\pm0.13$ \\
			& YZ (\(\mu\)m) & $4.56\pm0.70$ & $46$ & $8.41\pm0.70$ \\
			\hline
			& WD/WR (\(\mu\)m) & 36 & 10 & 400 \\
			& Design NA & 0.6 & 0.14 & 0.82 \\
			\hline
			\hline
			& $R$ (MHz) & $1.11\pm0.05$ & $1.33\pm0.07$ & $0.44$ \\
			CBG 1 & $\eta_{end-to-end}$ (\%) & $1.5\pm0.1$ & $1.8\pm0.1$ & $0.6$ & $1510$ \\
            & $\eta_{coll}$ (\%) & $9.0\pm1.3$ & $7.9\pm0.8$ & $15.4\pm2.9$  \\
			\hline
			& $R$ (MHz) & $1.66\pm0.10$ & & $0.90\pm0.14$ \\
			CBG 2 & $\eta_{end-to-end}$ (\%) & $2.2\pm0.1$ & N/A & $1.2\pm0.2$ & $1537$   \\
            & $\eta_{coll}$ (\%) & $13.5\pm2.1$ & & $31.7\pm7.9$ &  \\

			\hline
			& $R$ (MHz) & $1.09$ & $1.10\pm0.06$ & $0.76$ \\
			CBG 3 & $\eta_{end-to-end}$ (\%) & $1.4$ & $1.5\pm0.1$ & $1.0$ & $1490$ \\
            & $\eta_{coll}$ (\%) & $8.9\pm1.3$ & $6.6\pm0.7$ & $25.1\pm5.1$ \\
			\hline
			& $R$ (MHz) & $0.58$ & $0.35$ & $0.41$ \\
			CBG 4 & $\eta_{end-to-end}$ (\%) & $0.8$ & $0.5$ & $0.5$ & $1570$ \\
            & $\eta_{coll}$ (\%) & $4.7\pm0.7$ & $2.1\pm0.2$ & $14.4\pm2.7$ \\
			\hline 
			& & & &\\
			& $\eta_{setup}$ (\%) & $16.1\pm2.3$ &  $22.1\pm1.7$ &  $3.7\pm0.7$ \\
			& & & &\\
			
		\end{tabular}
		\caption{Statistics of the measurements taken with the lensed fiber, bare fiber, and microscope objective. For four CBGs, the count rate $R$, collection efficiency $\eta_{coll}$, end-to-end efficiency $\eta_{end-to-end}$, and wavelength are listed. The given values correspond to the arithmetic average and standard deviation from multiple measurement periods. If no standard deviation is noted only one measurement was performed.}
		\label{tab:Table1}
	\end{table*}
	\subsection{\label{sec:Brightness}Comparison of device efficiency}
    Apart from the spatial sensitivity, it is also of interest of investigate the maximum achievable count rate each configuration can provide. However, to quantify and make a reliable comparable study, we determine both the collection and end-to-end efficiency of each configuration. Here, the collection and end-to-end efficiency are defined as the fraction of the photons collected by the 3D-printed lens, the cleaved fiber end, and the microscope objective and fraction measured directly at the detectors.
    Since the setup transmission efficiency of the three configurations is different, it is important to determine the losses within each setup to evaluate the collection efficiency. The efficiencies can be calculated with $\eta_{coll}= R/\eta_{setup}f_{rep}$ and $\eta_{end-to-end} = R/f_{rep}$ with $R$, $f_{rep}$, and $\eta_{setup}$ being the measured count rate at the detectors, repetition rate of the excitation laser, and setup efficiency, respectively. For the following measurements, only non-resonant excitation was used.\\
    In Tab.\ref{tab:Table1} the count rate $R$, calculated collection efficiency, end-to-end efficiency of the four different CBGs, and the respective setup efficiency of each configuration are listed. The values with their respective uncertainty originate from multiple measurement repetitions and correspond to the arithmetic mean value and respective standard deviation. Here, the measured count rate $R$ is corrected for the imperfect second-order auto correlation function $g^{(2)}(0)$ with the factor $\sqrt{1-g^{(2)}(0)}$, resulting in the pure single-photon count rate \cite{g2corrected}. In the SI, the detailed spectra and $g^{(2)}(0)$ of each CBG are presented.

    Firstly, the measured count rate $R$ and $\eta_{end-to-end}$ is compared in each configuration. In Tab.\ref{tab:Table1}, no significant change in the count rate is observed for the measured CBGs. Thus, it is reasonable to state that $\eta_{end-to-end}$ is comparable for both configurations. When compararing these count rates to the microscope objective a drastic improvement is achieved for the fibers going from $\SI{0.44}{\mega\hertz}$ (CBG 1, microscope objective) to $\SI{1.33}{\mega\hertz}$ (CBG 1,  bare fiber) yielding a maximum factor of $3.0\pm0.2$. On average, the improvement is about $1.8\pm0.7$. The reason for the higher count rate is due to the enhanced $\eta_{setup}$ since no external optics are required for the fiber configurations. In our case, we were able to increase $\eta_{setup}$ from $\SI{3.7\pm0.7}{\percent}$ (microscope objective) to $\SI{22.1\pm1.7}{\percent}$ (bare fiber). It should be noted that $\eta_{setup}$ for the microscope objective in free space can generally be refined by optimizing the necessary optics possibly resulting in a sufficient $\eta_{setup}$ such that no advantage to the fibers can be gained. 
    \\
    In the next step, $\eta_{coll}$ is compared in each configuration. This value is important as it dictates how many photons can be collected into the first-lens and possibly be used in an experiment. After collecting the photons, the only loss sources are given by the setup, which as mentioned can generally always be improved. Here, the microscope objective seems to out-perform both fibers with $\SI{15.4\pm2.9}{\percent}$ (CBG 1, microscope objective) and $\SI{7.9\pm0.8}{\percent}$ (CBG 1, bare fiber). However, it is expected that $\eta_{coll}$ of the microscope objective and the lensed fiber should be in the same range since both their NAs are large enough to fully collect the single-photons emitted from the CBG (see SI). For the lensed fiber, the fiber in-coupling from the 3D-printed lens into the SMF-28 fiber is not accounted for in the measured $\eta_{setup}$ as it was not possible to experimentally determine it. Assuming that $\eta_{coll}$ of the lensed fiber and microscope objective should be about the same, the missing fiber in-coupling is estimated to be $\SI{42\pm12}{\percent}$.
    \\
    The fiber in-coupling could be improved by several optimization procedures. Firstly, reducing the shape deviations further by using more iterations for shape optimization could yield better lens performance. Using even more precise measurement methods, such as wavefront measurement \cite{ZhaoWangSiegleGiessen}, is a viable option to further enhance the lens quality. Finally, optimizing the length of the no-core fiber piece can allow for a better mode match from the lens to the fiber core. With these optimizations, the end-to-end efficiency of the lensed fiber could possibly exceed that of the bare fiber. In principle, we were able to show that the fibers can drastically enhance the measured count rate. This could greatly benefit experiments where multi-fold coincidences are required since here the coincidence rate scales exponential with the detected counts \cite{gaal2022near}. In the future, UHNA fibers could be used on both lensed and bare fiber to gain better fiber in-coupling since the smaller MFD should improve the mode matching from the CBG into the fiber \cite{rickert2023high}.


	\section{Conclusion}
    We performed measurements on four different CBGs in configurations: lensed fiber, bare fiber, and as a reference a microscope objective. Three aspects of the performance of these configurations were investigated: (i) imaging the sample surface demonstrated that the lensed fiber can reliably resolve the sample surface. (ii) The spatial sensitivity in the lateral XY- and vertical Z-direction of the lensed and bare fiber are within a couple of microns, with additional oscillations on the count rate when changing the fiber-sample distance with the bare fiber. Finally, (iii) the measured count rate with both fibers is equal and can out-perform the microscope objective of up to a factor of $3.0\pm0.2$. 
    \\
    As a conclusion we have shown that the lensed and bare fiber can be operated in an open and flexible approach, offering an advantage in terms of efficiency to a conventional microscope objective. Our measurements also present a systematic study of the spatial sensitivity of these fiber structures which is important for the realization of plug-and-play operation, improving further the scalability of single-photon sources. These findings pave the way for the efficient integration of semiconductor quantum emitters with optical fibers as modular components \cite{Sartison2021, Weber2024} and shall contribute to the realization of the future fiber-based quantum network for secure communications where fiber-coupled single-photon sources in the telecom C-Band are crucial.

	\section*{Funding}
H.G., P.R. and S.J. acknowledge Baden-Wuerttemberg-Stiftung (Opterial), European Research Council (Advanced Grant Complexplas, PoC Grant 3DPrintedOptics), Bundesministerium für Bildung und Forschung (Printoptics, QR.X, QR.N), Deutsche Forschungsgemeinschaft (DFG, German Research Foundation) (SPP1839 Tailored Disorder, 431314977/GRK2642), University of Stuttgart (Terra Incognita), Gips-Schüle-Stiftung, Carl-Zeiss-Stiftung (EndoPrint3D, QPhoton), and Vector Stiftung (MINT Innovations). \\
P.M., S.L.P., M.J. and N.T. acknowledge the funding by the German Federal Ministry of Education and Research (BMBF) via Project QR.X (16KISQ013) and QR.N (16KIS2207). 
Additional funding was also provided via the project EQSOTIC. This project was funded within the QuantERA II Programme that has received funding from the EU’s H2020 research and innovation programme under the GA No 101017733, and with funding organisation BMBF (with project number 16KIS2060K). This work was also funded by the Deutsche Forschungsgemeinschaft (DFG, German Research Foundation, 469373712/GRK2642) and the Quantum Technology BW via Project TelecomSPS.

\section*{Acknowledgements}
The authors thank Sascha Kolatschek for his contrbiution in the FDTD simulations of the CBG far-field.

	\FloatBarrier
	\newpage
	\bibliographystyle{apsrev4-1}
	\bibliography{main}

\end{document}